# EFFECTS OF OX-LDL ON MACROPHAGES NAD(P)H AUTOFLUORESCENCE CHANGES BY TWO-PHOTON MICROSCOPY

*Ching-Ting Lin[1], En-Kuang Tien[1], Szu-Yuan Lee[1], Long-Sheng Lu[2], Chau-Chung Wu[3,4], Chen-Yuan Dong, Chii-Wann Lin[1]*

Institute of Biomedical Engineering, National Taiwan University[1]
Graduate Institute of Parmacology, National Taiwan Universiy, College of Medicine[2]
Departments of Internal Medicine and Primary Care Medicine, National Taiwan University, College of[3] Medicine. Department of Internal Medicine Eda Hospital, Kaoshiung[4]
Chen-Yaun Dong, Department of Physics, National Taiwan University[5]

**ABSTRACT**

Ox-LDL uptakes by macrophage play a critical role in the happening of atherosclerosis. Because of its low damage on observed cells and better signal-to- background ratio, two-photon excitation fluorescence microscopy is used to observe NAD(P)H autofluorescence of macrophage under difference cultured conditions— bare cover glass, coated with fibronectin or poly-D-lysine. The results show that the optimal condition is fibronectin coated surface, on which, macrophages profile can be clearly identified on NAD(P)H autofluorescence images collected by two-photon microscopy. Moreover, different morphology and intensities of autofluorescence under different conditions were observed as well. In the future, effects of ox-LDL on macrophages will be investigated by purposed system to research etiology of atherosclerosis.

## 1. INTRODUCTION

The interaction between leukocytes and the adhesion molecules results in the adhesion between monocytes and endothelium cells. These monocytes then move to sub-endothelial space and differentiate into resident macrophages. At the beginning of atherosclerosis, LDL and other nutrient macromolecules are transported through endothelial cells into the vascular intima. LDL will stay in the sub-endothelium because of charged interacting with matrix proteoglycans. These LDL is attacked and oxidized by reactive oxygen species, called oxidized-LDL (ox-LDL). Generally, LDL is recognized by the LDL receptors and absorbed by cells. However, the ox-LDL will change the structure of apolipoprotein B on LDL, and is recognized by the scavenger receptors of macrophage instead of LDL receptor. By pinocytosis, ox-LDL is engorged by macrophages and form macrophage foam cells, which will become activated leading to a state of chronic inflammation. [1]

In present research, fluorescence microscopy has been used to investigate oxidation of LDL [2], and ox-LDL uptake by macrophage was demonstrated [3]. However, physiologic changes of macrophages during "eating" ox-LDL are still obscure. NAD(P)H, which emit autofluorescence (in fact, NAD(P)H are main source of autofluorescence) and play important roles in cellular metabolism [4][5], can generally reflect metabolic conditions. A metabolic flux that increases NAD(P)H : $NAD^+(P)$ ratio will raise intensity of autofluorescence as well, for NAD(P)H has much greater fluorescence yield than $NAD^+(P)$ [4]. Nevertheless, traditional fluorescence microscopy uses UV light as light source and it can result in significant damage on observed cells. [4]

Since the first two-photon excitation fluorescence microscope was realized in 1990, two-photon microscopy had widespread application in biology and biomedicine research. [6] Compared with traditional optical microscopy, two-photon microscopy uses longer incident wavelength thus it has lower Rayleigh scattering. The incident wavelength of a two-photon microscope is usually between 700-1000nm. In this range biological tissues have low absorption and thus two-photon microscopy plays better preferment when observing thick bio-samples. Besides, unlike one-photon excitation, of which excitation probability is proportional to intensity of incident light, excitation probability of two-photon excitation is proportional to square of intensity of incident light. Therefore, two-photon microscopy has an important characteristic: only molecules around focal point of objective will emit fluorescence. Thus a two-photon microscope does not have to use pinholes to filer fluorescence comes out of focal point, which is the main principle of confocal microscopy. According to this property, two-photon microscopy has some more advantages such as high signal-to-background ratio, low photo bleaching and photo damage, naturally optical section, fluorescence and incident wavelength could be easily separated. [6].

Even though two-photon microscopy has worse resolution compared with one-photon microscopy because its longer incident wavelength, advantages of two-photon microscopy still make it an ideal tool to observe NAD(P)H and understanding metabolism of cells. Actually, it was used to measurement NAD(P)H changes in macrophage[7]. In this experiment, two-photon microscopy will be used to measure autofluorescence changes of macrophages during ox-LDL uptake according to the protocol in Fig.1.

Fig. 1 Research protocol

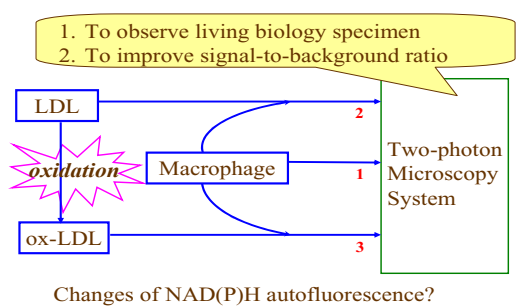

## 2. MATERIAL AND METHODS

A custom-made two-photon microscope is used to observe autofluorescence from macrophage described in the following sections.

### 2.1. Two-Photon Microscopy System

The two-photon microscopy system was moderated from an inverted microscope (Olympus IX71). Light source is a Ti-sapphire mode-lock pulse laser (Spectra-physics, Tsunami) with a pump laser (533nm, Spectra-physics, Millennia). Incident laser for observation was set to be 780nm and ~5mW on the specimen. The objective used in the experiments is 60x water objective, NA=1.2 (Olympus, UPlanApo). Gathered light is separated into three detector channels by dichoric and filter: channel 1: 400-495nm, channel 2: 495-560nm, channel 3: 560-680nm. Each channel obtains a 256 x 256 pixel, 16 bit unsigned image. NAD(P)H fluorescence is mainly collected by channel 2 and channel 4, since its emits spectra is from 400nm-500nm, and have maximum around 450nm [8]

### 2.2. Cell Culture and Specimen Preparation, Measurement

Macrophage (J774A.1, Bioresource Collection and Research Center) was cultured in chambered coverglass (Nunc Lab-Tek) with medium (L-DMEM with 10%FBS) for ~48hours, and then washed with PBS for two times and equilibrated in PBS with 25mM glucose. Autofluorescence was measured by two-photon microscopy system directly after 5 min equilibration in PBS. LDL and ox-LDL are obtained from National Taiwan University Hospital.

To confirm the condition of measurement, macrophage was cultured on chambered coverglass with three different coating types: bare cover glass, coated with fibronectin (10ug/ml) and coated with poly-D-lysine. Poly-D-lysine is a kind of synthetic compounds that alters surface charges on the culture substrate, and thus cell adhesion and protein absorption are enhanced. On the other hand, fibronectin is a widely distributed glycoprotein and it could be used as a substrate to promote adhesion of cells by its centra-binding domain RGD sequence.

Furthermore, to make sure the fluorescence collected by the system produced from metabolism of macrophage cells, intensity of autofluorescence is compared before and after addition of saturated potassium solution. Since we expected potassium solution would have effect on metabolism of cells in few minutes, high power of laser, 30mW and so on, is used in order to get apparent difference between control and experimental sets. In control set, macrophage were scanned rapidly for three minutes; and in experimental set, saturated potassium solution was added to cells after one minute scanning and then continued another two minutes scanning.

In the future, macrophages would be fed with different concentration of LDL and ox-LDL, and autofluorescence signals were obtained every period by two-photon microscopy system. To decide the period of every scanning, a test experiments should be done for fear of making observable damage to the cells.

### 2.3. Image Process and Statistics

IMAGEJ was used for RGB merger and text images acquirement, and to measure average intensity of pictures (the extremely bright region was excluded). MATLAB was used to obtain intensity-color mapping images.

## 3. PRESENT RESULT AND FUTURE WORK

Fig 2. (a), (b), and (c) are these images of cells cultured on bare cover glass, coated with fibronectin and poly-D-lysine, respectively. The scale is 20um and the color bar is from 0 to 64, indicates the intensity of fluorescence, demonstrates that NAD(P)H signal can be successfully detected by our custom-made two-photon microscopy.

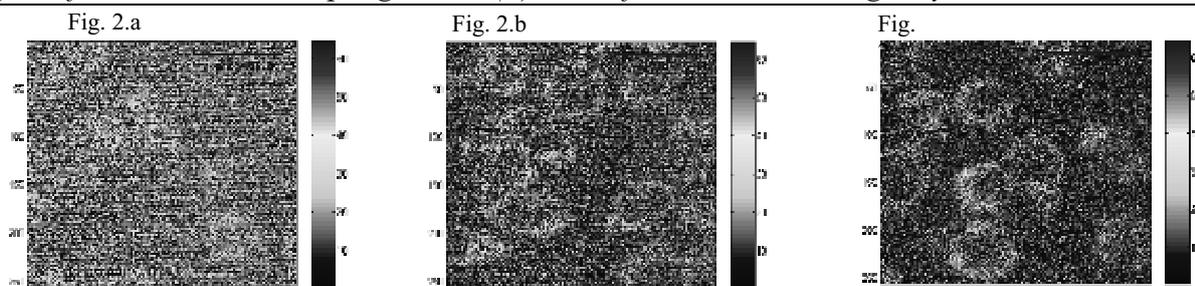

Fig 2. NAD(P)H autofluorescence image of macrophage J774A.1. Images are 16-bit unsigned, 256 x 256 pixels. Scale bar is 20um, color bar is from 0-64. Cells were cultured on chambered coverglass (a) without coating, (b) coated with fibronectin, (C) coated with poly-D-lysine.

In addition, under different adhesion status, the intensity of NAD(P)H autofluorescence signal is changed as well. The morphology of macrophages cultured on fibronectin coating surface comparing that on the other two substrates. The cells cultured on bare cover glass had blurred images thus it was difficult to identify cells. However, cells cultured on other two substrates had clear profile. On the other hand, cells on fibronectin surface had more uniform intensity than them on poly-D-lysine surface. It confirms that NAD(P)H signal can reflect condition of cells.

Besides, saturated potassium solution was used to destroy normal metabolism of macrophage, since whose ion concentration is much higher than it in cells. At the beginning, averaged intensity of autofluorescence of macrophage in control and experimental sets were 7.164 and 6.303, shown as Fig 3 (a) and (b). After three minutes rapidly scanning, averaged intensity of control set decreased to 6.303, showed 12.01% decay in intensity. In experimental set, saturated potassium solution resulted in 31.58% decay (from 6.675 to 4.567). The decrease in intensity in control set may result from high power of laser, but much larger range intensity decrease may because of saturated potassium solution. It suggested that autofluorescence collected is produced from metabolism of macrophage.

In conclusion, because its low damage on observed cells, two-photon microscopy could be a idea tool to observe NAD(P)H autofluorescence changes of macrophage, and reflect its metabolism. In the future work, to measure NAD(P)H signals of macrophages feed with different concentration of LDL (or ox-LDL) is a practice way to understand interaction between macrophage of LDL (or ox-LDL) and thus help investigating happening of atherosclerosis.

## 4. ACKNOWLEDGEMENT

This work is supported by THE MINISTRY OF COUNTRY for its program of research university and NANO CENTER FOR SCIENCE AND TECHNOLOGY.

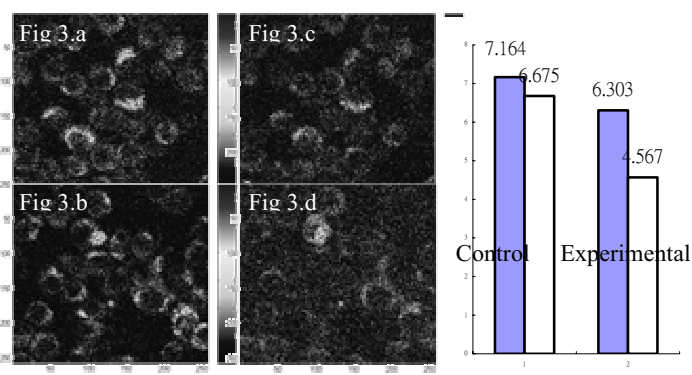

Fig 3(a) and (b) shows the beginning averaged intensity of auto-fluorescence of macrophage, in control and experimental sets. (c) and (d) shows them after three minutes rapidly scanning. Saturated potassium solution was added into experimental sets after two minutes scanning. There were 12.01% and 31.58% decrease in intensity between (a) and (c), (b) and (d), respectly.